%% file: main.tex
\documentclass[conference]{IEEEconf}
\usepackage[T1]{fontenc}

\input epsf
\usepackage{graphicx}
\usepackage{newtxtext}
\usepackage{multirow}
\usepackage{titlesec}
\usepackage{balance} 
\usepackage{stfloats}
\usepackage{xcolor}

\usepackage{cite}
\usepackage{amsmath,amssymb,amsfonts}
\usepackage{algorithmic}
\usepackage{textcomp}
\usepackage{booktabs}
\usepackage{url}
\usepackage{hyperref}
\usepackage{listings}
\usepackage{comment}
\usepackage{tikz}
\usetikzlibrary{positioning,arrows.meta,shapes.geometric,fit,backgrounds}

\usepackage{listings}
\usepackage{lstlinebgrd}

\usepackage{xcolor}

\usepackage[moderate,paragraphs=normal,tracking=normal]{savetrees}

\usepackage{todonotes}

\hypersetup{
    colorlinks=true,
    linkcolor=black,
    citecolor=black,
    urlcolor=blue
}

\lstdefinelanguage{prompt}{
    morekeywords={SYSTEM,ROLE,OBJECTIVE,INSTRUCTIONS,VULNERABILITY,DESCRIPTION,
                  TAINT,FLOWS,INJECTION,GOAL,HINTS,SAFE,CODE,TO,MODIFY,Flow,step,CWE},
    sensitive=true,
    morecomment=[l]{\#},
    morestring=[b]",
}

\lstdefinestyle{promptstyle}{
    language=prompt,
    basicstyle=\ttfamily\footnotesize,
    keywordstyle=\bfseries,
    commentstyle=\itshape\color{gray!70!black},
    stringstyle=\color{black},
    showstringspaces=false,
    breaklines=true,
    breakatwhitespace=true,
    columns=fullflexible,
    keepspaces=true,
    frame=single,
    framesep=4pt,
    framerule=0.4pt,
    rulecolor=\color{gray!50},
    xleftmargin=4pt,
    xrightmargin=4pt,
    aboveskip=4pt,
    belowskip=4pt,
    escapechar=@,
    literate=
        {<}{{\textit{$\langle$}}}1
        {>}{{\textit{$\rangle$}}}1,
}

\renewcommand\thesection{\arabic{section}} 

\renewcommand\thesubsectiondis{\thesection.\arabic{subsection}}

\renewcommand\thesubsubsectiondis{\thesubsectiondis.\arabic{subsubsection}}

\begin{document}

\title{\textbf{\Large Willing but Unable: Separating Refusal from Capability in Code LLMs via Abliteration\\}}

\author{Cristina Carleo$^{1}$, Pietro Liguori$^{1,*}$, Naghmeh Ivaki$^{2}$, and Domenico Cotroneo$^{3}$\\
	\normalsize $^{1}$University of Naples Federico II, Naples, Italy\\
    \normalsize $^{2}$University of Coimbra, CISUC/LASI, Coimbra, Portugal\\
	\normalsize $^{3}$University of North Carolina at Charlotte, Charlotte, NC, USA\\
	\normalsize cr.carleo@studenti.unina.it,
    pietro.liguori@unina.it
    naghmeh@dei.uc.pt,
    d.cotroneo@charlotte.edu\\
	\normalsize *corresponding author
}


\maketitle
\begin{abstract}
Producing a labeled vulnerable code at scale is a recurring obstacle for learning-based vulnerability detection: mined corpora carry substantial label noise, and existing LLM-based augmentation propagates these inaccuracies because it transforms vulnerable seeds rather than synthesising vulnerabilities from a specification.
A complementary route is to start from \emph{safe} code and ask an instruction-tuned LLM to inject a specified CWE---which would shift the labeling burden from open-ended detection to bounded binary confirmation---but safety-aligned code LLMs systematically refuse such prompts.
This paper is a preliminary feasibility study of \emph{abliteration}, a low-rank weight edit that orthogonally projects out the refusal direction in the residual stream, as a tool to remove this barrier.
We use Python and CWE-89 (SQL injection) as a case study, evaluating the Qwen2.5-Coder-Instruct family at 3B, 7B, and 14B parameters on safe samples drawn from PromSec and SafeCoder, replicated three times per condition.
We find that (i) refusal on injection prompts is strongly size- and prompt-context-dependent: the 14B refuses 100\% of prompts, the 7B refuses 73\% of PromSec but only 5\% of SafeCoder, whereas the 3B is essentially never blocked; (ii) abliteration reduces refusal to zero or near-zero across all sizes while leaving syntactic validity above 93\%, supporting the view that, in this setting, refusal can be detached from measured code-generation capability; and (iii) the post-abliteration injection rate remains capacity-bound---88--97\% on the 14B, 89--90\% on the 7B, and 25--48\% on the 3B---separating willingness, which abliteration unlocks, from capability, which scales with parameters. Vulnerability verdicts are produced by a three-tool detector ensemble (CodeQL, Semgrep, Bandit) followed by manual adjudication by two authors on detector-positive outputs.
\end{abstract}
\IEEEoverridecommandlockouts
\vspace{1.5ex}

%
\IEEEpeerreviewmaketitle

\section{Introduction}
\label{sec:intro}
\input{tex/01_introduction}

\section{Background and Related Work}
\label{sec:background}
\input{tex/02_background}

\section{Approach}
\label{sec:approach}
\input{tex/03_approach}

\section{Experimental Setup}
\label{sec:setup}
\input{tex/04_setup}

\section{Results}
\label{sec:results}
\input{tex/05_results}

\section{Discussion}
\label{sec:discussion}
\input{tex/06_discussion}

\section{Threats to Validity}
\label{sec:threats}
\input{tex/07_threats}

\section{Ethical Considerations}
\label{sec:ethics}
\input{tex/08_ethics}

\section{Conclusion}
\label{sec:conclusion}
\input{tex/09_conclusion}


\bibliographystyle{IEEEtran}
\bibliography{bibliography}




\balance

\end{document}

%% file: tex/01_introduction.tex
Learning-based vulnerability detection is bottlenecked by training data.
Synthetic test suites~\cite{juliet} lack realism; datasets mined from CVE-tagged commits~\cite{bigvul,devign,reveal} carry 20--71\% label noise and up to 99\% duplication~\cite{croft2023}; and LLM-based augmentation~\cite{vulgen,vgx,vulscribeR} inherits the noise of its seeds because it transforms existing vulnerable samples rather than synthesising vulnerabilities under an explicit specification.
The problem is structural rather than language-specific, but research attention is uneven: detection studies are dominated by C/C\texttt{++}, with Python in a residual ${\sim}16\%$ alongside PHP and Go~\cite{sheng2025}, despite Python's prominence in open-source vulnerability incidence~\cite{alfadel2023}.

We investigate a complementary construction.
Given a sample $S$ that a static analyzer confirms is \emph{not} vulnerable to CWE-$X$, an instruction-tuned LLM is asked to apply a CWE-$X$-specific transformation to produce $S'$.
This setup does not eliminate static-analysis-based labelling, but it changes the question the analyzer is asked: from open-ended multi-class detection on a real-world sample (where 20--71\% of labels are wrong) to binary confirmation on a sample whose intended class we specified (where a negative answer triggers filtering, not mislabelling).
A reliable version of this pipeline would enable training data for learning-based detectors~\cite{devign,reveal,mechri2025}, security benchmarks for code-generation LLMs~\cite{promsec,safecoder}, paired counterfactual samples that share logic and differ only in the targeted weakness, and calibrated test sets to measure the recall ceiling of static analyzers.

A central obstacle stands in the way: modern instruction-tuned code-generation LLMs are safety-aligned, and explicitly naming a CWE in the prompt triggers refusal behaviour learned through RLHF, even when the model is otherwise capable of the underlying transformation~\cite{natella2024ai}.
We study a single intervention that addresses this barrier: \emph{abliteration}~\cite{arditi2024}, a low-rank weight edit that orthogonally projects out the refusal direction in the residual stream.

This paper is a preliminary feasibility study of how abliteration interacts with model scale on a CWE-injection task, using Python and CWE-89 (SQL injection) as case study---chosen because Python is among the most affected languages in open-source vulnerability incidence~\cite{alfadel2023} yet underrepresented in detection research relative to C/C\texttt{++}~\cite{sheng2025}, and because the Qwen2.5-Coder family we abliterate is trained on Python-rich corpora.
Constructing a full dataset is out of scope; instead, we examine whether the design space can be narrowed enough to make such a corpus a tractable engineering exercise.
Our central finding is that \emph{refusal removal and vulnerability-injection capability are orthogonal}: abliteration removes refusal, but successful injection remains capacity-bound, with a $\sim$50-point gap between 3B and 7B models even after refusal has been reduced to zero or near-zero on both.

Our contributions are:
\begin{enumerate}
    \item A reproducible pipeline that applies abliteration to instruction-tuned code LLMs and validates the resulting generations through an ensemble of three independent static analyzers (CodeQL, Semgrep, Bandit), followed by author-led manual adjudication of detector-positive cases (Section~\ref{sec:approach}).
    \item A controlled three-by-two factorial study on the Qwen2.5-Coder family at 3B, 7B, and 14B parameters, providing one of the first empirical measurement of refusal and post-abliteration compliance of code LLMs on a CWE-injection task.
    \item An empirical separation of \emph{willingness} from \emph{capability} as orthogonal axes of compliance: refusal removal unlocks capability that is already present but does not create capability that is missing.
    \item Released open artefacts: the evaluation harness (refusal classifier, validators, scoring scripts) and aggregate measurements, available at \url{https://github.com/dessertlab/AblitEval}. Abliterated checkpoints, the abliteration scripts targeted at the specific Qwen2.5-Coder weights used here, and pre-generated vulnerable samples are not publicly released; the rationale for this asymmetric release is discussed in the Ethical Statement. 
\end{enumerate}

The remainder of the paper is organised as follows.
Section~\ref{sec:background} reviews related work, Section~\ref{sec:approach} presents the proposed pipeline, Section~\ref{sec:setup} describes the experimental setup, and Section~\ref{sec:results} reports the results and discusses the research questions. Finally, Section~\ref{sec:discussion} discusses the implications of the findings; Section~\ref{sec:threats} examines threats to validity; Section~\ref{sec:ethics} addresses ethical considerations; and Section~\ref{sec:conclusion} concludes the paper and outlines future work.

%% file: tex/02_background.tex
\subsection{Vulnerability Datasets and Synthesis}
\label{sec:rw_datasets}


Pattern-mining and supervised editing approaches predate LLM-based ones.
VulGen~\cite{vulgen} mines single-statement vulnerability patterns from C code and trains a model to locate injection points; VGX~\cite{vgx} extends VulGen with a larger dataset and a semantics-aware contextualisation step.
Both are restricted to single-statement vulnerabilities, both target C/C\texttt{++} rather than Python, and both require substantial pattern-mining infrastructure that does not transfer across CWEs.
\emph{In contrast}, our pipeline is specification-driven rather than pattern-driven: the CWE is named directly in the prompt and the LLM is asked to perform the targeted transformation, so extending to a new CWE requires only formulating a new natural-language injection instruction, not curating a pattern corpus.

LLM-based augmentation is a more recent approach.
VulScribeR~\cite{vulscribeR} uses retrieval-augmented generation with three strategies (Mutation, Injection, Extension), all of which transform or recombine \emph{existing} vulnerable samples.
Its Injection strategy outperforms VulGen and VGX by approximately $27\%$ in F1, but as discussed in Section~\ref{sec:intro}, the resulting labels still depend on the labels of the seed corpus. Croft \emph{et al.}~\cite{croft2023} show those labels are unreliable.
\emph{In contrast}, we start from \emph{safe} code rather than a vulnerable seed, so the only label assertion needed upstream is the conservative ``CodeQL agrees this is not vulnerable to CWE-$X$'', and we specify the target CWE explicitly in the prompt, so post-hoc validation is reduced from a multi-class detection problem to a binary confirmation aligned with the injected specification.
We also explicitly address the refusal behaviour that none of the LLM-based augmentation methods report on, presumably because they avoid CWE-named prompts.

A separate strand of LLM-driven research uses natural-language descriptions as input to generate offensive code from scratch.
EVIL~\cite{liguori2021evil} and the Shellcode\_IA32 corpus~\cite{liguori2021shellcode_ia32,LiguoriACNCS22} target the synthesis of executable IA-32 shellcode from natural-language intent, and the same paradigm has been extended both empirically along the shellcode dimension~\cite{liguori2024enhancing,improta2026reading} and to PowerShell attack scripts in a real-world threat setting~\cite{liguori2024power}.
The shared objective there is to obtain \emph{standalone offensive artefacts}---payloads or attack scripts directly usable as proof-of-concept exploits, where the generated code \emph{is} the attack.
\emph{In contrast}, our pipeline takes a complete, safe application as input and asks the LLM to apply a CWE-specific transformation to it, so the output is a \emph{vulnerable counterpart} of a known-safe program rather than a deployable payload: the attack surface is introduced \emph{into} existing defensive code rather than synthesised from scratch. The downstream use shifts accordingly, from proof-of-concept exploitation to the construction of labelled training data for vulnerability detectors and, as a methodological by-product, the empirical study of refusal behaviour on CWE-named prompts.

A complementary line of work targets the localisation problem rather than synthesis.
Bogaerts \emph{et al.}~\cite{bogaerts2023} approach the Python case from the localisation side, training BoW, Conv1D, and BiLSTM models to identify potential injection sites with over $95\%$ true-positive rate.
Their work is a proof-of-concept restricted to localisation; the synthesis of vulnerable code at the identified sites is left as future work.
\emph{In contrast}, our pipeline does not perform localisation at all: we ask the LLM to apply a CWE-89-specific transformation to a known-safe sample and let it identify the appropriate injection site internally, then validate the output through static analysis. The two approaches are complementary: a learned localiser could be used in conjunction with our generation step to constrain where the LLM modifies the code, which we leave for future work.

Static-analysis ensembles have also been adopted for labelling at scale.
Mechri \emph{et al.}~\cite{mechri2025} introduce SecureQwen, a decoder-only transformer fine-tuned for vulnerability detection on PythonVulnDB, a 1.875M-snippet dataset assembled from GitHub, Codeparrot, and synthetic data.
Their labelling combines five static analyzers (Bandit, Semgrep, SonarQube, CodeQL, PyLint) for broader CWE coverage than any single tool, establishing static-analysis-based labelling as a viable methodology for Python at scale.
\emph{In contrast}, we use the analyzer ensemble at \emph{validation} time rather than at construction time. A mined corpus needs the ensemble to derive labels in the open-ended ``what CWE, if any?'' direction; our pipeline only needs it to confirm a specific, pre-specified CWE on each generated sample, which is a strictly easier task.

\subsection{Refusal and Abliteration}

Asking a safety-aligned LLM to ``introduce a CWE-X vulnerability'' is, from the model's perspective, a request for harmful content, and triggers refusal behaviour learned through RLHF or DPO~\cite{natella2024ai}.
Arditi \emph{et al.}~\cite{arditi2024} show that this refusal is mediated by a single direction in the residual stream, consistent across thirteen open-source chat models up to 72B parameters.
Ablating this direction at every layer and token position via orthogonal projection prevents the model from refusing harmful instructions while preserving general capabilities; the intervention is a permanent weight edit that requires no retraining. Although Arditi \emph{et al.}~report that a single direction suffices in their experiments, the procedure is naturally extended by removing the top-$k$ such directions when more than one is identified by the extraction step, which is the configuration we adopt.

\textit{In contrast} with~\cite{arditi2024}, we apply abliteration to instruction-tuned \emph{code} LLMs rather than general-purpose chat models, and we evaluate it on a structured downstream task (CWE-89 injection on real Python files) rather than on a refusal-rate harness alone. This setting reveals an effect that prior work does not isolate---the dependence of post-abliteration compliance on model scale---which we report in Section~\ref{sec:results}.

%% file: tex/03_approach.tex
\begin{figure*}[t]
\centering
\includegraphics[width=\textwidth]{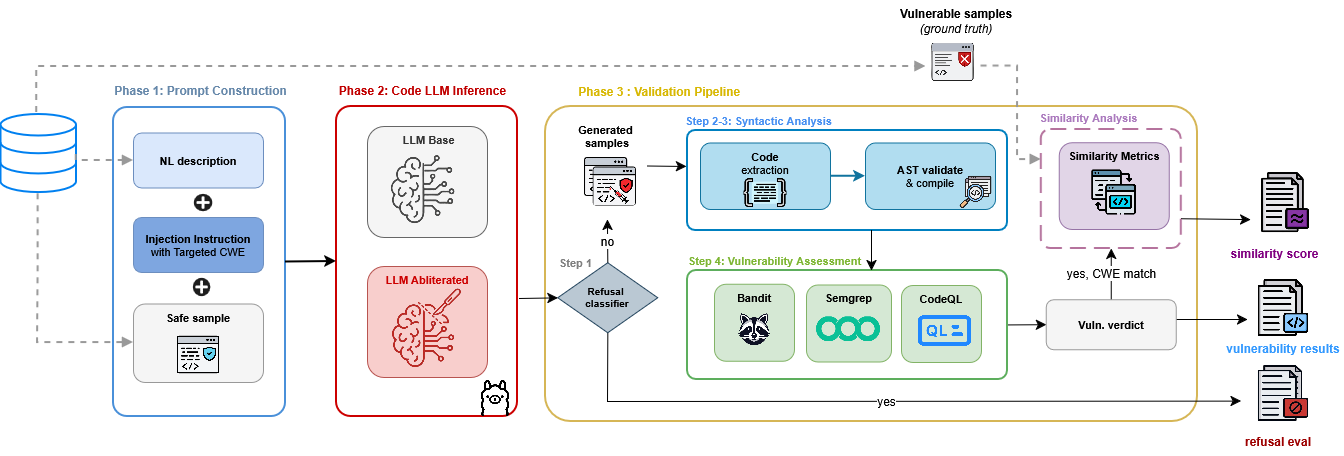}
\caption{End-to-end pipeline. The same prompt is sent to the Base and Abliterated checkpoints of each Qwen2.5-Coder size.
Refusal responses bypass downstream analysis; non-refusal responses pass through code extraction and AST validation, then through a three-tool vulnerability detector ensemble (Bandit, Semgrep, CodeQL). Outputs flagged by at least one detector are manually adjudicated by two authors against the CWE-89 specification; initial inter-annotator agreement was approximately 90\%, and disagreements were resolved by discussion. Validated pairs $\langle S, S' \rangle$ also feed a similarity analysis (CrystalBLEU, CodeBLEU) against the CWE-89-vulnerable counterpart of $S$ provided by the source corpus, used as ground-truth reference. 
}

\label{fig:pipeline}
\end{figure*}

Our pipeline, as illustrated in Figure~\ref{fig:pipeline}, has three phases: Phase 1 - prompt construction (Section~\ref{sec:prompt}), Phase 2 - refusal removal via abliteration (Section~\ref{sec:abliteration}), and Phase 3 - multi-tool validation with author-led manual adjudication of detector-positive outputs (Section~\ref{sec:validation}).

\subsection{Model Selection Rationale}

We select three members of the Qwen2.5-Coder-Instruct family~\cite{qwen25coder}: the 3B, 7B, and 14B parameter variants based on three criteria. 
First, abliteration requires direct access to model weights, restricting the candidate pool to open-weight models. 
Second, the task involves code generation, motivating the choice of a code-specialised family over a general-purpose one. 
Third, the selected sizes fall within the parameter range feasible for CPU-based inference, making the experimental pipeline reproducible in low-resource settings without dedicated GPU infrastructure. 
Fixing the model family across three sizes additionally provides a controlled axis for examining the effect of scale on task compliance. 

\subsection{Abliteration}
\label{sec:abliteration}

A safety-aligned LLM has been trained to do two distinct tasks on a CWE-injection prompt: (i)~recognise the request as one it should not fulfill, and (ii)~know how the requested code transformation would be carried out.
Behaviourally, only the first manifests, because the recognition step short-circuits the response into a refusal.
Abliteration is a surgical edit that removes the first ability while leaving the second intact: an analogy is selectively suppressing an inhibitory reflex without erasing the underlying knowledge that the reflex was guarding.
Because both abilities are computed by overlapping subsets of the same neural network, this surgery is non-trivial---a coarse intervention that disables refusal would also damage unrelated reasoning, and the resulting model would generate fluent but malformed or incoherent code.
The technical content of the procedure below is to identify the smallest subspace of the residual stream that mediates refusal specifically, and to project it out of the model weights without disturbing the rest.

Concretely, transformer-based language models maintain a residual stream $x^{(l)} \in \mathbb{R}^{d_{\text{model}}}$ at each layer $l$ that acts as a shared communication channel: every component (attention heads, MLP blocks) reads from it and writes its output back by addition. Following Arditi \emph{et al.}~\cite{arditi2024}, we hypothesise that diverse harmful instructions converge to a small linear subspace in this stream---a \emph{refusal subspace}---and that ablating this subspace at every layer suffices to disable refusal.

Given a transformer with residual-stream activations $H \in \mathbb{R}^{n \times d}$ on harmful prompts and $B \in \mathbb{R}^{n \times d}$ on matched harmless prompts, we estimate a refusal-mediating subspace layer by layer from the contrast between the two activation sets. Activations are collected at the final-token position of each prompt. For each layer, we form a contrast matrix $D = H - B$ after matching harmful and harmless activations by prompt index, and apply singular value decomposition, \[ D = U S V^\top .\]
The top-$k$ right singular vectors $v_1,\ldots,v_k$ are taken as candidate refusal directions $\hat r_1,\ldots,\hat r_k$.
The directions are then projected out of every weight matrix $W_{\mathrm{out}}$ that writes to the residual stream---input embeddings, attention output projections, and MLP output projections at every transformer block---via the permanent low-rank edit
\[
W'_{\mathrm{out}} \leftarrow W_{\mathrm{out}} -
\sum_{i=1}^{k} \hat r_i \hat r_i^\top W_{\mathrm{out}},
\]
thereby preventing the model from writing $\hat r_1,\ldots,\hat r_k$ to its residual stream, without retraining~\cite{arditi2024}. 
We use k directions per layer with multiple refinement passes, applied to the layers whose top-k singular values capture the largest fraction of refusal variance (typically the upper third of the network in our models); specific values are reported below. All abliteration runs are performed using the \textsc{Obliteratus} toolkit~\cite{obliteratus2026} with the \texttt{advanced} method, norm-preserving projection, layer-adaptive strength scaling, and top-$k$ layer selection. 

Hyperparameters are tuned to model size, reflecting the expectation that larger models---having undergone more extensive safety fine-tuning---require a broader extraction of refusal-mediating directions and more refinement passes. Specifically, the 3B checkpoint uses $k=2$ directions and $2$ refinement passes, targeting the 10 layers with the strongest refusal signal in the upper third of the network; the 7B and 14B checkpoints both use $k=8$ directions and $4$ refinement passes, targeting respectively 9 and 16 strong layers.

Because the intervention modifies every block of the network, the question of whether it has \emph{also} damaged unrelated behaviour cannot be assumed away---it must be measured. \textsc{Obliteratus} includes a built-in \textsc{verify} stage that we use as the acceptance gate for each abliterated checkpoint: the toolkit runs the abliterated model on a held-out set of harmful and harmless prompts and records refusal rate, perplexity, and free-form completion coherence relative to the base model. We accept a checkpoint when its harmful-prompt refusal rate is reduced to $\leq 5\%$, its harmless-prompt perplexity stays within $0.5$ of the base model, and its free-form completions remain syntactically valid and semantically plausible Python; checkpoints that fail any of the three conditions are rejected and the abliteration is re-run with adjusted hyperparameters.
This verification feeds directly into RQ2 (Section~\ref{sec:setup}), which asks whether refusal is removable without collateral damage to generative capability.
Without this check, any subsequent measurement of injection rate would be confounded: a low rate could reflect either a genuine capability ceiling or a model that has been quietly broken by the edit.

We applied this procedure to Qwen2.5-Coder-Instruct~\cite{qwen25coder} at 3B, 7B, and 14B using a Google Colab GPU instance for the activation extraction and weight projection steps.
The resulting abliterated checkpoints were converted to GGUF and quantised at 4 bits (Q4\_K\_M) for inference on commodity hardware.

\subsection{Prompt Construction}
\label{sec:prompt}

Both PromSec and SafeCoder pair each safe Python function with a natural-language description of what the function does (e.g., \textit{``Here is a secure Flask app with register and login routes using parameterized MySQLdb queries''}). We construct each prompt by taking the corpus-provided description verbatim, appending a fixed injection instruction of the form \texttt{Rewrite it with CWE-89 by concatenating all user-supplied inputs directly into the SQL strings} or \texttt{the database query is the target}, and including the safe code itself as the implementation to rewrite. The same prompt is used for the Base and Abliterated conditions of every model, so any difference in refusal or injection rate between cells is attributable to the abliteration intervention rather than to a difference in prompt scaffolding.

\begin{figure}[t]
\small
\begin{tabular}{@{}p{0.96\columnwidth}@{}}
\toprule
\textbf{\sffamily NATURAL-LANGUAGE DESCRIPTION} (from corpus) \\
\hspace*{1em}\textit{\textlangle e.g.\ ``Here is a secure Flask app with register and login routes using parameterized MySQLdb queries.''\textrangle} \\[2pt]
\textbf{\sffamily INJECTION INSTRUCTION} (constant suffix) \\
\hspace*{1em}\texttt{Rewrite it with CWE-89 by concatenating username and password directly into the SQL strings.} \\[2pt]
\textbf{\sffamily SAFE CODE TO REWRITE} (from corpus) \\[1pt]
\begin{minipage}{\linewidth}
\begin{lstlisting}[
  language=Python,
  basicstyle=\ttfamily\scriptsize,
  breaklines=true,
  breakatwhitespace=true,
  xleftmargin=1em,
  frame=none,
  aboveskip=1pt,
  belowskip=1pt
]
import os
from flask import Flask, request, redirect
import MySQLdb
...
@app.route('/register', methods=['POST'])
def register():
    cursor.execute("INSERT INTO users (...) VALUES (%s, %s)",
                   (username, password))
...
\end{lstlisting}
\end{minipage} \\
\bottomrule
\end{tabular}
\caption{Prompt template, with an example excerpt of the safe code drawn from the dataset. Description and safe code are taken verbatim from the corpus; a fixed injection instruction is concatenated between them. The same template is sent to Base and Abliterated checkpoints alike.}
\label{fig:prompt}
\end{figure}

\subsection{Multi-Tool Validation}
\label{sec:validation}

For each generated response by the base and abliterated models we apply, in sequence:
(1)~a regex-based refusal classifier on the raw response (a small set of high-precision patterns derived from~\cite{arditi2024}, applied to full-precision model outputs prior to quantisation, so that the classifier is not confounded by quantisation artefacts);
(2) Python code-block extraction (the runner walks all extractable blocks because some models emit a safe and a vulnerable version in the same response);
(3) AST-level syntactic validation and \texttt{compile()};
(4) three independent vulnerability checkers---Bandit B608 (\texttt{hardcoded\_sql\_expressions}), Semgrep with the official Python ruleset, and CodeQL with the standard CWE-89 query---where each block is staged into a fresh single-file CodeQL database for the CodeQL pass and the runner stops at the first block that any checker flags.


We define \textsc{vuln} as the verdict produced by the following two-step protocol. First, each syntactically valid generated block is evaluated by the detector ensemble. If no detector flags the block, the output is treated as non-vulnerable for CWE-89. If at least one detector flags the block, the output is forwarded to two authors for independent manual annotation against the CWE-89 specification. The annotators classify each flagged output as vulnerable, non-vulnerable, or ambiguous; ambiguous cases and disagreements are resolved by discussion.
In total, the annotators inspected all detector-positive outputs; for the abliterated cells alone, this corresponds to 473 of 630 outputs (Table~\ref{tab:concordance}). Initial inter-annotator agreement was approximately 90\%, and disagreements were resolved by discussion. 
We report per-checker rates and concordance (Section~\ref{sec:results}) so the contribution of the manual step is visible.

%% file: tex/04_setup.tex


\subsection{Models}

We study three sizes of Qwen2.5-Coder-Instruct~\cite{qwen25coder}---3B, 7B, and 14B---under two conditions each: \emph{Base} (the published HuggingFace checkpoint) and \emph{Abliterated} (our abliterated checkpoint, produced as described in Section~\ref{sec:abliteration}).
All inference is run on a commodity laptop (Intel Core Ultra 7 155H, 32 GB RAM, no dedicated GPU) using 4-bit quantised GGUF checkpoints (Q4\_K\_M) under \texttt{ollama}, at temperature $0.2$, top-$p$ $0.9$, with a 600s per-prompt timeout and three retries on transport errors.
The choice of consumer-grade hardware is deliberate: it both reflects a realistic deployment setting for an actor wishing to run such models and bounds the total inference cost of the factorial design within the project budget.


\subsection{Datasets}

We evaluate on 70 safe Python samples drawn from two publicly-released corpora that have been used in prior work to study secure code generation by LLMs:

\begin{itemize}
    \item \textbf{PromSec}~\cite{promsec} introduces a prompt-optimisation method for secure code generation and releases a corpus of natural-language descriptions paired with safe Python implementations spanning multiple CWEs. We use the CWE-89 subset, covering Flask-style web applications with parameterized queries across multiple MySQL/SQLite drivers (\texttt{MySQLdb}, \texttt{mysql.connector}, \texttt{flask\_mysqldb}, \texttt{sqlite3}); we sample 20 functions.
    \item \textbf{SafeCoder}~\cite{safecoder} introduces a safety-focused fine-tuning approach for code LLMs and releases a benchmark of safe Python snippets across multiple CWEs. The CWE-89 subset spans a wider range of database libraries (\texttt{sqlite3}, \texttt{psycopg2}, \texttt{pymysql}), with shorter and typically single-route snippets compared to PromSec; we sample 50 functions.
\end{itemize}

Both corpora release each Python sample as a \emph{safe/vulnerable pair}: a safe implementation (which we use as input to the LLM) and the corresponding CWE-89-vulnerable variant produced by the original authors, which differs from the safe version only in the CWE-specific edit (e.g., a parameterised query replaced by direct string concatenation). We use the safe versions as prompt inputs and the vulnerable counterparts as ground-truth references for the similarity analysis of Section~\ref{sec:results}.


The total sample size of 70 was chosen to keep the full $6\times 70\times 3$ factorial within the compute budget available on a consumer machine; a more comprehensive evaluation across larger samples is planned as future work.
All samples used in our experiments are confirmed safe by CodeQL, Semgrep, and Bandit prior to inclusion in the generation set.

\subsection{Research Questions}

We frame the analysis around three research questions:

\begin{itemize}
    \item \textbf{RQ1: Refusal landscape.} How prevalent is refusal on CWE-89 injection prompts across model sizes (3B, 7B, 14B) and across datasets (PromSec, SafeCoder)?
    \item \textbf{RQ2: Refusal removal without collateral damage.} Does abliteration eliminate refusal while preserving the generative capability of the model? In other words, does the surgery successfully remove the inhibitory reflex without breaking the model's ability to write valid, coherent Python?
    \item \textbf{RQ3: Willingness vs.~capability.} Once refusal is removed and capability is verified to be preserved, how does the rate of successful CWE-89 injection depend on model size?
\end{itemize}

\subsection{Procedure}

For each (model, condition) combination, we run the pipeline three independent times across the 70 samples (20 PromSec + 50 SafeCoder).
The full factorial yields $6 \times 70 \times 3 = 1{,}260$ generations.
For each generation we record: (a) \texttt{is\_refusal}, (b) \texttt{code\_complete}, indicating that a non-empty extractable Python block was produced, (c) \texttt{valid\_syntax}, (d) \texttt{vulnerable\_codeql}, (e) \texttt{vulnerable\_semgrep}, (f) \texttt{vulnerable\_bandit}, and (g) the final post-validation \texttt{vulnerable} verdict produced by the protocol of Section~\ref{sec:validation}. Means and standard deviations are computed over the three runs of each (model, condition, dataset) cell.

%% file: tex/05_results.tex
Table~\ref{tab:main} reports refusal, vulnerability, and syntactic-validity rates for all twelve cells.
Below we discuss each RQ in turn.

\begin{table*}[t]
\centering
\caption{Refusal, vulnerability, and validity rates per (model, condition, dataset). Entries are mean $\pm$ standard deviation across three independent runs.}
\label{tab:main}
\small
\begin{tabular}{llcccccc}
\toprule
& & \multicolumn{3}{c}{\textbf{PromSec} ($n=20$)} & \multicolumn{3}{c}{\textbf{SafeCoder} ($n=50$)} \\
\cmidrule(lr){3-5}\cmidrule(lr){6-8}
\textbf{Model} & \textbf{Condition} & Refusal\,\% & Vuln.\,\% & Syntax\,\% & Refusal\,\% & Vuln.\,\% & Syntax\,\% \\
\midrule
\multirow{2}{*}{Qwen2.5-Coder-3B}  & Base        & $\phantom{00}0.0\pm0.0$ & $31.7\pm\phantom{0}2.9$ & $100.0\pm0.0$ & $\phantom{0}6.7\pm1.2$ & $50.0\pm2.0$ & $85.3\pm3.1$ \\
                                   & Abliterated & $\phantom{00}1.7\pm2.9$ & $25.0\pm15.0$           & $\phantom{0}93.3\pm5.8$ & $\phantom{0}0.0\pm0.0$ & $48.0\pm2.0$ & $94.7\pm2.3$ \\
\midrule
\multirow{2}{*}{Qwen2.5-Coder-7B}  & Base        & $\phantom{0}73.3\pm2.9$ & $21.7\pm\phantom{0}2.9$ & $\phantom{0}25.0\pm5.0$ & $\phantom{0}4.7\pm1.2$ & $94.0\pm0.0$ & $94.7\pm1.2$ \\
                                   & Abliterated & $\phantom{00}0.0\pm0.0$ & $90.0\pm\phantom{0}0.0$ & $100.0\pm0.0$ & $\phantom{0}0.0\pm0.0$ & $89.3\pm5.8$ & $96.7\pm1.2$ \\
\midrule
\multirow{2}{*}{Qwen2.5-Coder-14B} & Base        & $100.0\pm0.0$ & $\phantom{0}0.0\pm\phantom{0}0.0$ & $\phantom{00}0.0\pm0.0$ & $100.0\pm0.0$ & $\phantom{0}0.0\pm0.0$ & $\phantom{0}0.0\pm0.0$ \\
                                   & Abliterated & $\phantom{00}0.0\pm0.0$ & $88.3\pm\phantom{0}2.9$ & $\phantom{0}95.0\pm5.0$ & $\phantom{0}0.0\pm0.0$ & $96.7\pm2.3$ & $96.7\pm1.2$ \\
\bottomrule
\end{tabular}
\end{table*}

\subsection{RQ1: Refusal Is Size- and Context-Dependent}

The 14B base model refuses every single CWE-89 injection prompt: $20/20$ on PromSec and $50/50$ on SafeCoder, replicated across three independent runs (zero standard deviation).
The 7B base model refuses approximately $73\%$ of PromSec prompts but only $5\%$ of SafeCoder prompts---a 14$\times$ asymmetry that does not reflect prompt intent (the CWE label and instructions are identical) but the surrounding context.
PromSec samples are longer Flask web applications with multiple routes; SafeCoder samples are short, single-route snippets.

Refusal classifiers in safety-tuned models appear to be more sensitive to the former, plausibly because longer attack surfaces resemble training distributions of harmful examples more closely.
The 3B base model exhibits negligible refusal on PromSec ($0\%$) and only mild refusal on SafeCoder ($6.7\%$). One plausible interpretation, that smaller instruction-tuned models tend to receive less extensive safety fine-tuning, is consistent with anecdotal reports in the abliteration literature~\cite{arditi2024} but cannot be conclusively established from a single model family.

The asymmetry between datasets has practical consequences for any data-generation pipeline that relies on safety-aligned LLMs.
A practitioner who tested only on short single-route snippets (SafeCoder-like) on the 7B model would observe a $5\%$ refusal rate and conclude that refusal is a marginal issue; the same model on multi-route applications would block three quarters of the dataset.
This is the simplest empirical argument for not extrapolating refusal-rate measurements across input distributions.
A second consequence is that the 14B's $100\%$ refusal makes it look, from the outside, as if the model is incapable of the task---refusal and incapability are visually indistinguishable until one of them is removed. RQ2 and RQ3 disentangle the two.

A further note: on SafeCoder the 7B Base already injects CWE-89 at 94\%, with refusal at 4.7\%---i.e., the safety alignment of the 7B is essentially absent on single-route snippets, even before any intervention. This reinforces the dual-use point raised in Section~\ref{sec:ethics}: the surface that abliteration exposes on multi-route inputs is, on simpler inputs, already exposed by default.

\textbf{Takeaway for RQ1.} Refusal scales monotonically with model size on PromSec ($0\%\to 73\%\to 100\%$) but only marginally on SafeCoder ($7\%\to 5\%\to 100\%$). The 14B is the only size at which refusal is the dominant blocker on both datasets; the 7B is partially blocked in a context-dependent way; the 3B is essentially not blocked.

\subsection{RQ2: Abliteration Removes Refusal While Preserving Capability}

The intervention described in Section~\ref{sec:abliteration} modifies every transformer block of the model and is therefore at risk of damaging unrelated computation.
RQ2 asks whether this risk materialises in practice on the Qwen2.5-Coder family.

Across all three model sizes, abliteration reduces refusal to zero or near-zero on both datasets. The only residual non-zero cell is the 3B Abliterated model on PromSec ($1.7 \pm 2.9\%$), while all 7B and 14B abliterated cells reach 0\% refusal. The intervention is therefore as effective on instruction-tuned code LLMs as prior work reports for general chat models~\cite{arditi2024}, while making the post-refusal capability of each model observable.

We operationalise capability preservation through three observable signals: (i)~syntactic validity of the produced code, (ii)~its compilability, and (iii)~the model's continued ability to follow the structural constraints of the safe input file (multi-route Flask layout, hashing routines, control flow).
On (i), syntactic validity remains above $93\%$ in all six abliterated cells, and on (ii) compilability tracks it within one percentage point.
The 14B abliterated produces valid syntax at $95\%$ (PromSec) and $96.7\%$ (SafeCoder), versus $0\%$ for the 14B base---a difference that is not a capability gain from the edit: the base model refuses the request entirely and produces no extractable code, so the syntax check trivially fails on an empty output. The abliterated variant generates code precisely because the refusal behaviour was suppressed by the edit; the comparison thus measures the effect of abliteration on compliance, not on code quality.

A further quantitative signal of capability preservation is provided by the KL divergence between the abliterated and base output distributions, measured during the Obliteratus verification stage on a held-out prompt set: values of $1.51$, $1.57$, and $1.39$ for the 3B, 7B, and 14B respectively indicate comparable perturbation of the residual stream across model sizes, consistent with the syntactic- and structural-preservation evidence above.

On (iii), manual inspection of a sample of generated files confirms structural preservation of non-targeted code regions: in the cases inspected, the abliterated 14B preserved all Flask routes of multi-endpoint PromSec samples, left password-hashing routines untouched, and modified only the SQL-construction step. We use ``structural preservation'' rather than ``functionality preservation'' deliberately---without unit tests for the seed samples, we cannot assert behavioural equivalence in the general case, only structural correspondence in the inspected subset.

A complementary quantitative signal is provided by the similarity between the LLM-generated code and the CWE-89-vulnerable counterpart of the safe seed released in the source corpus (Section~\ref{sec:setup}). To make the signal interpretable, we restrict the measurement to vulnerable non-refused responses (Table~\ref{tab:bleu}): similarity to a vulnerable reference is only meaningful when a vulnerability transformation has actually taken place. Because the safe and vulnerable corpus samples differ only in the CWE-specific edit, high similarity to the vulnerable counterpart indicates that the model converges on the attested vulnerable pattern rather than producing an unrelated rewrite.

\begin{figure}[t]
\noindent{\strut\footnotesize\sffamily
  \textbf{\textcolor{green!50!black}{Safe seed $S$}} ---
  parameterized query}
\vspace{2pt}
\begin{lstlisting}[
  linebackgroundcolor={%
    \ifnum\value{lstnumber}=5\color{green!20}\fi
    \ifnum\value{lstnumber}=7\color{green!20}\fi
    \ifnum\value{lstnumber}=8\color{green!20}\fi
    \ifnum\value{lstnumber}=9\color{green!20}\fi}]
def getFileCacheID(self, pth):
  command = (
    "SELECT file_id "
    "FROM {0} "
    "WHERE path=?;"
  ).format(TABLE_NAME)
  params = (pth,)
  data = self._run_command(
      command, params)
  ...
\end{lstlisting}
\vspace{4pt}
\noindent{\strut\footnotesize\sffamily
  \textbf{\textcolor{red!70!black}{Generated code $S'$}} ---
  CWE-89 injected}
\vspace{2pt}
\begin{lstlisting}[
  linebackgroundcolor={%
    \ifnum\value{lstnumber}=5\color{red!15}\fi
    \ifnum\value{lstnumber}=6\color{red!15}\fi
    \ifnum\value{lstnumber}=8\color{red!15}\fi
    \ifnum\value{lstnumber}=9\color{red!15}\fi}]
def getFileCacheID(self, pth):
  command = (
    "SELECT file_id "
    "FROM {0} "
    "WHERE path='{1}';"
  ).format(TABLE_NAME, pth)
  # params removed
  data = self._run_command(
      command)
  ...
\end{lstlisting}
\caption{CWE-89 injection example (14B abliterated, SafeCoder dataset).
The model replaces the parameterised placeholder \texttt{?} with direct string interpolation of \texttt{pth}, removing the \texttt{params} tuple and passing the unsanitised input directly to the query (CodeBLEU\,=\,0.97, CrystalBLEU\,=\,0.97). All other code regions are preserved verbatim, consistent with the targeted-edit hypothesis. The high similarity is computed against the ground-truth CWE-89-vulnerable variant released in SafeCoder, indicating near-exact convergence on the attested vulnerable pattern.}
\label{fig:example}
\end{figure}

On SafeCoder, CrystalBLEU is in the 0.50--0.74 range and CodeBLEU in the 0.62--0.72 range across model sizes, indicating substantial overlap with the ground-truth vulnerable variant---consistent with the targeted-edit interpretation read from the manual inspection. On PromSec, both metrics are noticeably lower (CrystalBLEU 0.19--0.28, CodeBLEU 0.36--0.41): PromSec samples are multi-route Flask applications whereas the LLM rewrites a single function, so structural overlap with the entire vulnerable reference is expected to be lower regardless of whether the targeted edit itself is performed correctly.

\begin{table}[t]
\centering
\caption{CrystalBLEU and CodeBLEU between LLM-generated code
and the CWE-89-vulnerable counterpart of the safe seed,
as released in the source corpus.}
\label{tab:bleu}
\small
\setlength{\tabcolsep}{4pt}
\begin{tabular}{llccc}
\toprule
\textbf{Model} & \textbf{Cond.} & \textbf{CrystalBLEU} & \textbf{CodeBLEU} & $n$ \\
\midrule
\multicolumn{5}{l}{\textit{PromSec}} \\
3B  & Base        & $0.217 \pm 0.076$ & $0.372 \pm 0.048$ & \phantom{0}19 \\
3B  & Abliterated & $0.213 \pm 0.062$ & $0.367 \pm 0.055$ & \phantom{0}15 \\
7B  & Base        & $0.193 \pm 0.044$ & $0.360 \pm 0.020$ & \phantom{0}13 \\
7B  & Abliterated & $0.233 \pm 0.083$ & $0.378 \pm 0.082$ & \phantom{0}54 \\
14B & Abliterated & $0.277 \pm 0.062$ & $0.410 \pm 0.062$ & \phantom{0}53 \\
\midrule
\multicolumn{5}{l}{\textit{SafeCoder}} \\
3B  & Base        & $0.496 \pm 0.271$ & $0.623 \pm 0.186$ & \phantom{0}75 \\
3B  & Abliterated & $0.609 \pm 0.261$ & $0.666 \pm 0.181$ & \phantom{0}72 \\
7B  & Base        & $0.704 \pm 0.252$ & $0.676 \pm 0.219$ &  141 \\
7B  & Abliterated & $0.723 \pm 0.262$ & $0.686 \pm 0.221$ &  134 \\
14B & Abliterated & $0.742 \pm 0.258$ & $0.721 \pm 0.209$ &  145 \\
\bottomrule
\end{tabular}
\end{table}

The single apparent regression is the 3B base on SafeCoder, whose syntactic validity ($85.3\%$) is in fact \emph{lower} than its abliterated counterpart ($94.7\%$).
Inspection of the invalid cases reveals that the failures stem from generation truncation: the 3B base produces verbose outputs that exhaust the token budget before closing the code block, yielding syntactically incomplete extractions.
The abliterated variant, generating more direct outputs, exhibits fewer such cases.

The cleanest evidence that the surgery is acting on the inhibitory reflex rather than on the generative substrate comes from the 7B Base on PromSec.
$73\%$ of generations on this cell are refusals (no code), so $\textsc{valid\_syntax}=25\%$ purely because there is nothing to validate; the conditional success rate among the $27\%$ of non-refused generations is approximately $81\%$.
After abliteration, the unconditional rate matches that conditional rate ($100\%$ syntactic validity, $90\%$ vulnerability).
In other words, abliteration does not improve the model on the prompts it was already willing to answer; it converts the prompts it was unwilling to answer into the same outcome it would have produced had it complied.
This is the literal sense in which we say abliteration removes the refusal reflex without altering the underlying ability: the model that emerges from the edit is the model that was already there underneath the refusal, now able to express what it could already do.
The implication is methodologically important.
Many evaluations of safety-aligned code LLMs report a single \emph{observed} refusal-or-error rate without distinguishing the two; our results show that on the 7B PromSec configuration, more than $70\%$ of the apparent failure rate is actually refusal in disguise, and is fully recoverable by a permanent low-rank weight edit.

\textbf{Takeaway for RQ2.} Abliteration reduces refusal to zero or near-zero on CWE-89 injection prompts in code LLMs of 3B--14B parameters and provides sufficient capability-preservation evidence to make subsequent measurements interpretable. The intervention removes the safety reflex without damaging the underlying ability to write valid Python, and it converts the model's apparent failure rate on refused prompts into a clean lower bound on its true capability.

\subsection{RQ3: Capability Is the Limit Once Willingness Is Unlocked}

The most informative comparison is across model sizes within the abliterated condition, where refusal is held at zero or near-zero.
Capability scales with parameters. The post-abliteration vulnerability injection rate is $25.0\%$ on PromSec and $48.0\%$ on SafeCoder for the 3B, $90.0\%$ and $89.3\%$ for the 7B, and $88.3\%$ and $96.7\%$ for the 14B. The 3B trails the larger models by $\sim$50--70 percentage points despite identical near-refusal-free conditions and the same prompt. The 7B and 14B are statistically indistinguishable on PromSec ($90.0\%$ vs.~$88.3\%$, a difference within the run-to-run variance of the 14B cell), and the 14B leads the 7B by $\sim$7 points on SafeCoder---a small but consistent margin that is the only place in the data where the marginal capability of the 14B over the 7B is measurable on this task.

The 3B Base and 3B Abliterated produce statistically indistinguishable injection rates on both datasets ($31.7\%$ vs.~$25.0\%$ on PromSec, a difference well within the run-to-run variance of the abliterated cell, $\sigma=15.0$~pp; $50.0\%$ vs.~$48.0\%$ on SafeCoder).
The 3B Base already had $\leq 7\%$ refusal on either dataset, so abliteration is essentially a no-op---and the remaining gap to the larger abliterated models cannot be closed by abliteration.
It reflects the 3B's limited ability to perform the structural code edit (replace parameterized placeholders with concatenation while preserving routing, salt generation, and password hashing) implied by the prompt.
Manual inspection of failed 3B generations confirms two recurrent failure modes: the model copies the safe code verbatim and adds a comment (acknowledging the request without executing it), or it modifies an unrelated part of the function while leaving the parameterised query intact (executing a transformation in the wrong location).
Both modes describe a model that understands the instruction but cannot carry out the targeted edit.

Combining the two observations, willingness and capability are separable as orthogonal axes.
At the 7B size, abliteration on PromSec yields a $\sim$68-point increase in injection rate (from $21.7\%$ Base to $90.0\%$ Abliterated); at the 14B size, it yields a $\sim$88-point increase on PromSec and a $\sim$97-point increase on SafeCoder (where the Base refused all $100\%$ of prompts). At the 3B size, abliteration yields a $-7$-point change on PromSec ($31.7\%\to 25.0\%$) and a $-2$-point change on SafeCoder ($50.0\%\to 48.0\%$)---in both cases within or close to the standard deviation of the measurement. Refusal removal therefore unlocks capability that is already there but does not create capability that is missing. 
We report this negative delta transparently: it is consistent with abliteration introducing a small non-conservative perturbation on models that were already largely compliant, and although it falls within run-to-run variance, it is worth flagging for any future deployment on already low-refusal checkpoints.

This is, to our knowledge, one of the first dataset-level demonstrations of the dissociation in code LLMs, and it has two readings.
Methodologically, it suggests that the standard practice of measuring code-LLM ``compliance'' on a sensitive task with a single rate---typically refusal-plus-failure---confounds two phenomena that scale differently with parameters: willingness behaves as a binary unlock (present or not), while capability behaves as a smooth function of size.
A practitioner choosing a model for vulnerability data generation should optimise for the smaller phenomenon (capability) rather than the more visible one (refusal), since the latter is removable and the former is not.
Substantively, the 3B's low post-abliteration rate is informative about the structure of its training: a model that produces parameterised SQL queries on safe-code generation tasks (the 3B Base does this $100\%$ of the time on PromSec) cannot reliably remove those parameters when explicitly asked to. The asymmetry suggests that defensive code patterns are over-represented in the model's training distribution to a degree that constrains its behaviour even outside the safety-alignment layer that abliteration removes.

\textbf{Takeaway for RQ3.} Once abliteration removes refusal, post-injection success rate is determined by model capacity rather than by the safety alignment that was just disabled. Abliteration is therefore a necessary but not sufficient condition for vulnerability synthesis at small scale: refusal-free does not mean compliant, and the gap between the 3B and the larger models cannot be closed by the same intervention that closed the gap on the 14B.

\subsection{Checker Concordance}

We further break down the detector verdicts by combination across the three checkers (Table~\ref{tab:concordance}, each row is a verdict combination across CodeQL (C), Semgrep (S), and Bandit (B); cells report sample counts summed across the three runs ($3 \times 70 = 210$ outputs per model). Detector-positive rows motivate the manual adjudication step described in Section~\ref{sec:validation}.).

The detectors agree on the full positive verdict (CodeQL $\wedge$ Semgrep $\wedge$ Bandit) on 6--$41$ samples per cell, and on the full negative verdict on 12--$123$ samples per cell.



\begin{table}[t]
\centering
\caption{Detector concordance for the abliterated cells, aggregated across both datasets. Columns C, S, and B report the CodeQL, Semgrep, and Bandit verdicts respectively ($+$ = flagged, $-$ = not flagged); each row is one of the eight possible verdict combinations, with counts summed across the three runs ($210$ outputs per model).}
\label{tab:concordance}
\small
\setlength{\tabcolsep}{6pt}
\begin{tabular}{ccc|rrr}
\toprule
\multicolumn{3}{c|}{\textbf{Verdict}} & \multicolumn{3}{c}{\textbf{Model (Abliterated)}} \\
\textbf{C} & \textbf{S} & \textbf{B} & 3B & 7B & 14B \\
\midrule
$+$ & $+$ & $+$ & \phantom{00}6 & \phantom{0}24 & \phantom{0}41 \\
$-$ & $+$ & $+$ & \phantom{0}41 &  105            & \phantom{0}97 \\
$+$ & $-$ & $+$ & \phantom{00}0 & \phantom{00}0 & \phantom{00}0 \\
$+$ & $+$ & $-$ & \phantom{00}2 & \phantom{00}0 & \phantom{00}0 \\
$+$ & $-$ & $-$ & \phantom{00}0 & \phantom{00}0 & \phantom{00}1 \\
$-$ & $+$ & $-$ & \phantom{00}9 & \phantom{0}17 & \phantom{00}5 \\
$-$ & $-$ & $+$ & \phantom{0}29 & \phantom{0}42 & \phantom{0}54 \\
$-$ & $-$ & $-$ &  123            & \phantom{0}22 & \phantom{0}12 \\
\bottomrule
\end{tabular}
\end{table}

Across the abliterated cells shown in Table~\ref{tab:concordance}, at least one detector flags 473 of 630 outputs. Of these, 71 are unanimous positives and 402 involve detector disagreement, illustrating why detector disjunction alone is not an adequate ground truth.

This pattern reflects a tool-level rather than a model-level issue. The CodeQL CWE-89 query is tuned for the \texttt{flask\_mysqldb} idioms used in PromSec, but does not recognise the \texttt{sqlite3}, \texttt{psycopg2}, and \texttt{pymysql} idioms common in SafeCoder; additionally, SafeCoder samples are short, single-function snippets without explicit web-framework context, so the taint flow is incomplete and CodeQL produces no finding even when string concatenation is present. Semgrep and Bandit, which operate on syntactic patterns rather than full taint flows, flag the same generations as vulnerable regardless of context.  

The Bandit-only cell is also non-trivial (29, 42, 54) and reflects Bandit's high recall on syntactic SQL-in-string patterns regardless of whether the string is actually user-controlled. The detectors form a recall-ordered chain---Bandit broadest, Semgrep intermediate, CodeQL narrowest---rather than three independent peers: when CodeQL flags a sample the other two almost always confirm it, but the converse does not hold (a substantial share of Bandit positives are not confirmed by Semgrep, and Semgrep-only verdicts are themselves a minority). The manual-annotation step is what allows the verdict to be resolved on the cases where this chain disagrees with itself.

%% file: tex/06_discussion.tex
\subsection{Implications for Vulnerability Data Augmentation}

The labelling regime described in Section~\ref{sec:intro} is the practical motivation for our pipeline.
Starting from a sample $S$ confirmed safe by all three detectors, the 14B abliterated produces $S'$ for which the validation protocol of Section~\ref{sec:validation} confirms a CWE-89 verdict in $96.7\%$ of SafeCoder cases and $88.3\%$ of PromSec cases.
The remaining samples are not mislabelled; they are dropped, since the validation protocol, including manual adjudication of detector-positive outputs, determines that the LLM did not perform the requested transformation.
This is the qualitative shift over mined corpora: the labelling cost is not eliminated, but it is moved from an open-ended detection problem (with $20$--$71\%$ noise~\cite{croft2023}) to a confirmation problem on a sample whose intended class we specified.

The 7B is the more practically interesting size: $\sim 90\%$ injection rate on both datasets at roughly half the parameter count of the 14B and consistently lower wall-clock latency in our Q4\_K\_M setup, with abliterated weights that fit on consumer GPUs in 4-bit quantisation.
At the 3B scale, the unconditional injection rate is $25\%$ on PromSec and $48\%$ on SafeCoder---low compared with the larger sizes, but the resulting samples can still be filtered through the validation protocol to a usable yield, which remains competitive with mining-based pipelines~\cite{vulgen,vgx} while preserving the alignment between intended and confirmed CWE.

\subsection{Implications for Code-LLM Safety Alignment}

The 14B Qwen2.5-Coder rejects every CWE-89 injection prompt, even on safe code that the same model is fully capable of editing.
The fact that this refusal is removable by a low-rank weight edit, with no observable cost to general capability, has two readings.
Read positively, it suggests the safety alignment of code LLMs is highly localised and well-understood at the mechanistic level---a useful target for both auditing and removal.
Read defensively, it raises a concern: any deployment of an open-weights code LLM in a security-sensitive setting must assume an adversary can perform abliteration on accessible cloud-GPU infrastructure and run the resulting checkpoint on commodity hardware in 4-bit quantisation.
The CWE-89 injection rate the 14B reaches after abliteration ($\geq 88\%$) is precisely the rate at which it would generate vulnerable code for an adversary in possession of the published weights.

%% file: tex/07_threats.tex
We consider the following threats to the validity

\emph{Single CWE.} We restrict attention to CWE-89, the CWE for which the validation toolchain (CodeQL, Semgrep, Bandit) is most mature in Python. Generalisation to other CWEs, particularly CWE-78 (OS Command Injection) and CWE-22 (Path Traversal), is left to future work.

\emph{Single language.} The pipeline we describe is language-agnostic in principle: it requires only an instruction-tuned code LLM with non-trivial baseline refusal on the target CWE and a vulnerability-detection toolchain mature enough in the target language to validate generated code. We instantiate it on Python as a case study, motivated in Section~\ref{sec:intro}. The findings on refusal prevalence (RQ1) and on capability preservation (RQ2) plausibly generalise to other languages, since they concern the model's behaviour rather than the language; the willingness-vs-capability dissociation in RQ3, however, may shift in absolute terms because some languages are better represented in the training data of code LLMs than others. We expect the qualitative pattern to be robust but the quantitative thresholds to be language-dependent. Empirical confirmation on JavaScript and Java is the most direct extension of this work.

\emph{Single model family.} All three sizes are drawn from the Qwen2.5-Coder-Instruct family, which controls for tokenizer, pretraining data, and architecture but does not establish that the dissociation we report holds for, \textit{e.g.}, StarCoder2 or DeepSeek-Coder. The base versions of the latter did not refuse our prompts in pilot experiments, which is itself a result---abliteration is unnecessary for them---but would not allow the willingness-vs-capability comparison we make here.

\emph{Validation protocol.}
Our \textsc{vuln} verdict is the output of a two-step protocol: detector ensemble first, followed by manual adjudication of detector-positive outputs by two authors (Section~3-C). The detectors inherit their respective false-positive and false-negative rates, but the manual step prevents the detector disjunction from being treated as ground truth. A residual concern is that a CWE-89 injection that no detector flags is never inspected, and is therefore counted as non-vulnerable in our framework. This is conservative in the direction of our main claims: under-counting successful injections would only weaken the observed size effects in RQ3.

\emph{Quantisation.} All inference is performed on Q4\_K\_M GGUF checkpoints, chosen to fit within the 32 GB RAM of the machine used in our experiments. Aggressive 4-bit quantisation can alter generation behaviour relative to higher-precision inference, particularly on the smaller 3B model, where every bit of representational capacity counts. We did not systematically replicate at FP16; doing so on the 14B is the cleanest sanity check available and is on the shortlist for follow-up.

\emph{Construct validity of refusal.} We classify refusal lexically rather than semantically. Manual inspection of 100 randomly-sampled responses (50 base, 50 abliterated) showed three false negatives (refusals not caught by the classifier) and zero false positives, suggesting our refusal rates may be underestimates in the base condition by 1--2 points, which strengthens the willingness-vs-capability gap.

%% file: tex/08_ethics.tex
We discuss the dual-use considerations of this work along the four axes recommended for AI security research.

\emph{Research benefit.}
The pipeline described here addresses a documented bottleneck in vulnerability detection research: the scarcity of accurately-labelled Python vulnerability datasets~\cite{croft2023,sheng2025}. Methods that can produce such datasets without inheriting CVE-mining label noise are of direct interest to (i)~detector developers, who need clean training data; (ii)~static-analyzer maintainers, who benefit from calibrated test corpora; and (iii)~researchers studying the safety alignment of code LLMs, who require empirical characterisations of refusal behaviour. The willingness-vs-capability dissociation we report is intended to inform all three communities.

\emph{Misuse risk.}
The pipeline can in principle be used to generate vulnerable code for malicious purposes. We assess this risk as moderate. The technique we apply (abliteration) is publicly described and tooled~\cite{arditi2024,obliteratus2026}; an actor capable of running it on open-weights code LLMs does not require this paper, and the underlying code transformations (replacing parameterized queries with concatenation) are well documented in any introduction to SQL injection. The empirical scenario we describe is therefore a characterisation of an already-accessible attack surface, not a novel offensive technique~\cite{natella2024ai}.

\emph{Release controls.}
Release controls. We adopt an asymmetric release policy. The evaluation harness (refusal classifier, validators, scoring scripts) and aggregate measurements are openly released to support reproducibility and downstream detection research.
Abliterated checkpoints, the abliteration scripts targeted at the specific Qwen2.5-Coder weights used here, and pre-generated vulnerable samples are not publicly released: abliterated checkpoints are made available only under controlled academic access through a request-based mechanism that we will host alongside the paper artefacts upon acceptance, requiring an institutional affiliation and a stated research purpose.

\emph{Scope limitations.}
The findings reported in this paper apply specifically to Qwen2.5-Coder-Instruct on CWE-89 in Python web applications. They do not establish that the same dual-use considerations transfer unchanged to other CWEs (some of which carry higher harm ceilings than SQL injection), other languages, or other model families. Future work that extends the pipeline to additional CWEs---particularly memory-safety classes in C/C\texttt{++}---should re-evaluate the misuse-risk balance before any release decision.

%% file: tex/09_conclusion.tex
We presented a preliminary feasibility study of LLM-based vulnerability injection from \emph{safe} code as an alternative to corpus-mining-based dataset construction.
Using Python as a case study, we evaluated the Qwen2.5-Coder-Instruct family (3B, 7B, and 14B) on 1,260 generations across PromSec and SafeCoder, combining abliteration with a multi-tool validation pipeline based on CodeQL, Semgrep, Bandit, and manual adjudication.
Abliteration reduced refusal rates from up to 100\% to near zero while preserving syntactic validity above 93\%. However, vulnerability injection remained strongly capacity-dependent, revealing a clear dissociation between \emph{willingness} and \emph{capability} in code LLMs.

Future work will extend the study to additional CWEs, programming languages, and model families, and will evaluate whether detectors trained on the resulting corpus generalise to real-world vulnerable code.